\documentclass[11pt,twoside]{article}
\usepackage{cspm2015}

\aspvoltitle{Coimbra Solar Physics Meeting 2015: Ground-based Solar Observations in the Space Instrumentation Era}
\aspvolauthor{I. Dorotovi\u{c}, C. Fischer, and M. Temmer., eds.}

\resetcounters

\def\figspath{.} 

\begin{document}

\title{Observations and modelling of Helium lines in solar flares}

\author{Paulo J.~A. Sim\~oes, Lyndsay Fletcher, Nicolas Labrosse, and Graham S. Kerr \\
\affil{SUPA School of Physics and Astronomy, University of Glasgow, Glasgow, G12 8QQ, UK; \email{paulo.simoes@glasgow.ac.uk}}}

\paperauthor{Sim\~oes, P.~J.~A.}{paulo.simoes@glasgow.ac.uk}{}{University of Glasgow}{SUPA School of Physics and Astronomy}{Glasgow}{}{G12 8QQ}{UK}
\paperauthor{Fletcher, L.}{lyndsay.fletcher@glasgow.ac.uk}{}{University of Glasgow}{SUPA School of Physics and Astronomy}{Glasgow}{}{G12 8QQ}{UK}
\paperauthor{Labrosse, N.}{nicolas.labrosse@glasgow.ac.uk}{}{University of Glasgow}{SUPA School of Physics and Astronomy}{Glasgow}{}{G12 8QQ}{UK}
\paperauthor{Kerr, G.~S.}{g.kerr.2@research.gla.ac.uk}{}{University of Glasgow}{SUPA School of Physics and Astronomy}{Glasgow}{}{G12 8QQ}{UK}

\begin{abstract}
We explore the response of the He {\sc ii} 304 \AA\ and He {\sc i} 584 \AA\ line intensities to electron beam heating in solar flares using radiative hydrodynamic simulations. Comparing different electron beams parameters, we found that the intensities of both He lines are very sensitive to the energy flux deposited in the chromosphere, or more specifically to the heating rate, with He {\sc ii} 304 \AA\ being more sensitive to the heating than He {\sc i} 584 \AA. Therefore, the He line ratio increases for larger heating rates in the chromosphere. A similar trend is found in observations, using SDO/EVE He irradiance ratios and estimates of the electron beam energy rate obtained from hard X-ray data. From the simulations, we also found that spectral index of the electrons can affect the He ratio but a similar effect was not found in the observations. 
\end{abstract}

\section{Introduction}
The formation of He lines in the solar chromosphere and transition region has been subject of intense investigation. Three main mechanisms have been proposed for the formation of He {\sc ii} lines:
collisional excitation, photoionisation and radiative recombination \citep[e.g.][]{1975MNRAS.170..429J,1976ApJ...207L.199A}, and it has been discussed that ultimately all three may have a influence during flares, although each process may have a different levels of importance for the different He lines. The He {\sc ii} 304 \AA\ line is the most intense line in the solar spectrum shortward of 1000 \AA, and thus it has a relatively important role on the energy balance through radiative cooling, especially during flares. While quiet Sun observations of He lines have been abundant \citep[e.g.][]{1978ApJ...220..683M,1978ApJ...222..707G}, flare observations are scarce, mainly limited by the transient nature of flares and small FOV of past spectrometers such as OSO-7 \citep[e.g.][]{1976ApJ...203..509L} or \emph{Skylab} \citep[e.g.][]{1989SoPh..120..309P,1992ApJ...386..364L}. As such, most of our knowledge of the evolution of He in flares comes mostly from modelling efforts.

The most straightforward comparison between observations and models is the comparison of line intensity ratios, as they can provide insights into the effects of the competing excitation mechanisms.  Typically, models are limited to one-dimensional grids to describe the height structure of the atmosphere so the flare area cannot be taken into account. On the other hand, recent systematic Sun-as-a-star observations by the Extreme ultraviolet Varibility Experiment \citep[EVE, ][]{2012SoPh..275..115W}, on board of the Solar Dynamics Observatory (SDO), provided the line irradiances in the range 50--1050 \AA\ at moderate spectral resolution (1 \AA) for a large number of flares. Under the reasonable assumption that the emitting areas of the different He lines are essentially the same at a given instant for a given flare, the irradiance ratios can be directly compared to intensity ratios obtained from models. For specific events, SDO/AIA 304 \AA\ images and SDO/EVE spectra can be combined to estimate the emitting area and recover the average line intensities. However, the complex three dimensional structure of real flaring atmospheres can not be reproduced by the current models. 

In this paper, we present results of flare simulations of He emission and compare with flare observations from EVE. More specifically, we investigate the relative intensity of the lines He {\sc ii} 304 \AA\ and He {\sc i} 584 \AA\ for different electron beams to their response to the energy deposition in the chromosphere. 

\section{Modelling He lines in flares}

We employed radiative hydrodynamic (RHD) simulations using the RADYN code \citep{1995ApJ...440L..29C,1997ApJ...481..500C}, modified for flare modelling with heating supplied by a beam of electrons and also accounting for X-ray and extreme ultraviolet (EUV) backwarming from the corona \citep{2005ApJ...630..573A}. The code solves the equation of hydrodynamics, radiative transfer and statistical equilibrium in a one-dimensional atmosphere, treated in non-LTE for transitions of H, Ca and He. Detailed infomation about the model can be found elsewhere \citep{1999ApJ...521..906A,2005ApJ...630..573A,2015ApJ...809..104A}, and it has been used extensively recently \citep{2010ApJ...711..185C,2012MSAIS..19..117R,2015A&A...578A..72K,2015ApJ...813..125K,2015SoPh..tmp...61K}.

In this work, we focus on the emission lines He {\sc ii} 304 \AA\ and He {\sc i} 584 \AA\ for two main reasons: 1) these are the two strongest He lines observed systematically by SDO/EVE, providing the means for an statistical investigation and 2) both lines are calculated in full detail by RADYN. Observationally, both lines are also free of strong blends of other lines, with the major contribution being from a weak Si {\sc xi} line on the red wing of He {\sc ii} 304 \AA.

Starting from the same initial atmosphere, we simulated 4 scenarios changing only the spectral index $\delta$ of the electron beam: $\delta=$ 5, 6, 7, 8. The beam was injected with a triangular time profile lasting 60 seconds, with a maximum energy flux of $F=10^{10}$ erg s$^{-1}$ cm$^{-2}$ at 30 seconds. The low energy cutoff was set at $E_c=20$ keV. Hereafter we refer to these models as F10. For the same atmosphere, the low energy cutoff affects the height at which most of the electrons stop collisionally, while the spectral index affects how the energy deposition is spread out in the chromosphere: with a harder $\delta$ (i.e. smaller values) the electrons can penetrate deeper into the denser layers, while with a softer $\delta$ (i.e. larger values) most of the electrons will deposit their energy at a much thinner layer. The immediate consequence is that the maximum heating rate $Q$ in the latter case will be higher than the former -- for the same low energy cutoff $E_c$ and energy flux $F$ \citep[see Figure 5 in][]{2015ApJ...809..104A}. A typical RADYN run is very demanding computationally, so in order to explore the cases with different values for the energy flux $F$, we took advantage of two other runs made available elsewhere\footnote{Kowalski, A.~F. and Allred, J.~C., IRIS-4 Workshop, Boulder, CO, USA, 2015}: $10^9$ erg s$^{-1}$ cm$^{-2}$ and $10^{11}$ erg s$^{-1}$ cm$^{-2}$, models F9 and F11 hereafter. For these latter cases, a beam with $E_c=25$ keV and $\delta=4.5$ and a constant energy flux $F$ was injected for a given time (20 seconds for F11 and 200 seconds for F9). 

For each simulation, the He {\sc ii} and He {\sc i} line profiles were integrated over wavelength at each second of solar time providing the time evolution of the line intensity, and these are shown in Figure \ref{fig1}, along with the ratios $R=\lambda$304/$\lambda$584, for F10.

\articlefigure{\figspath/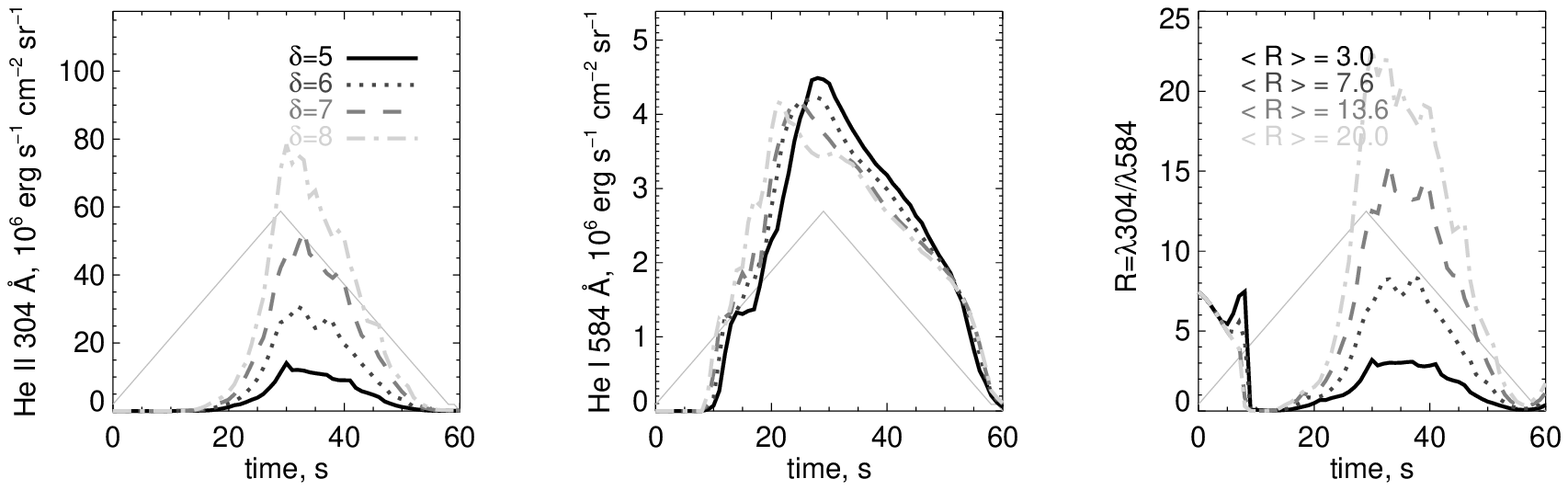}{fig1}{He {\sc ii} 304 \AA\ and He {\sc i} 584 \AA\ intensities calculated from the four F10 models, and their intensity ratio $R=\lambda$304/$\lambda$584. The grey triangular curves indicate the time profile of the energy injection.}

From Figure \ref{fig1} we see that the maximum of He {\sc ii} intensity varies by a factor of $\approx$8 for different $\delta$, while the He {\sc i} intensity does not change much in these models (less than 7\% at peak intensities). The ratio $R$ obtained from the F10 models varies from $\approx$3 (for $\delta=5$) to $\approx$20 (for $\delta=8$) at the peak of the emission. For the four F10 models, the maximum heating rate is on the order of $Q \approx 200 $ erg s$^{-1}$ cm$^{-3}$ during most of the duration of the simulated flares. These results indicate that the spectral index $\delta$ has a direct consequence for the intensity of the He {\sc ii} line.

The heating rate $Q$ is also controlled by the total amount of energy input into the model (i.e. energy flux). We verified the effect of the energy flux using models F9 and F11. For these models, despite the energy flux input being constant the line intensities do present time variations due to the dynamic response of the chromosphere to the beam heating. These changes in the atmosphere, specially changes in the column depth, also affect how it continues to be collisionally heated. The average heating rate in F9 is within the range $Q = 10 \sim 30$ erg s$^{-1}$ cm$^{-3}$, with the ratio $R$ between $1.4 \sim 2.1$ while the average heating rate in F11 is within the range $Q = 5 \sim 20 \times 10^3$ erg s$^{-1}$ cm$^{-3}$, with the ratio $R$ between 6 -- 20. Although $R$ varies during a particular simulation, the maximum values of $R$ are well associated with heating rates $Q$. Thus, the combined results from the models F9, F10, and F11 confirm the relationship of the heating rate and the ratio $R=\lambda$304/$\lambda$584: stronger heating leads to larger $R$ ratios. 

\section{SDO/EVE observations}

We now proceed to compare the above results from simulations with He {\sc ii} and He {\sc i} flare data obtained by SDO/EVE\footnote{EVL data, version 5.}. Although the heating rate cannot be obtained directly from observations, the electron beam parameters, namely the electron energy rate $E_\mathrm{rate}$, $E_c$, and $\delta$), can be estimated from hard X-ray (HXR) data under the assumption of the cold collisional thick-target model, using standard Solar Software (SSW) routines \citep[see e.g.][ and references therein]{2015SoPh..tmp...62S,2015ApJ...813..125K}. We selected four events observed by EVE with a good signal-to-noise ratio for both He {\sc ii} and He {\sc i} irradiances and HXR data available from either RHESSI \citep{2002SoPh..210....3L} or Fermi Gamma-ray Burst Monitor \citep[GBM, ][]{2009ApJ...702..791M}. In order to obtain the electron energy flux (erg s$^{-1}$ cm$^{-2}$) it is necessary to estimate the area of the HXR sources, which is not available for the flares observed by Fermi. We assume that the HXR footpoint areas of each flare have similar sizes, and hence we use the $E_\mathrm{rate}$ as an estimate of the energy content being deposited in the chromosphere. The He {\sc ii} 304 \AA, He {\sc i} \AA\ flare excess irradiances, their ratio $R$, spectral index $\delta$ and the energy rate contained by the electrons are shown in Figure \ref{figobs} for the events X2.2 SOL2011-02-15, M9.3 SOL2011-08-04, M9.3 SOL2011-07-30 and M9.9 SOL2014-01-01.

\articlefigurefour{\figspath/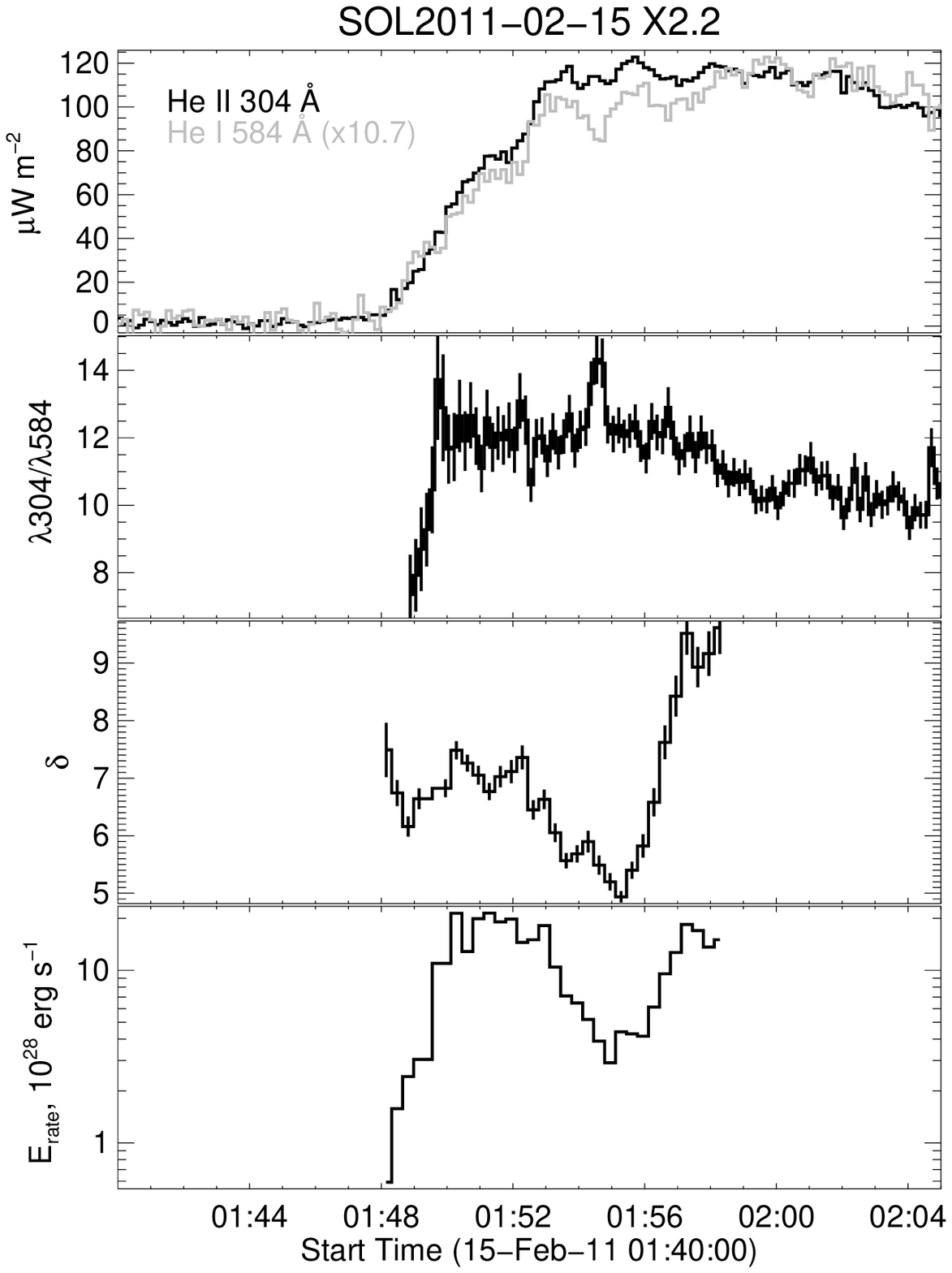}{\figspath/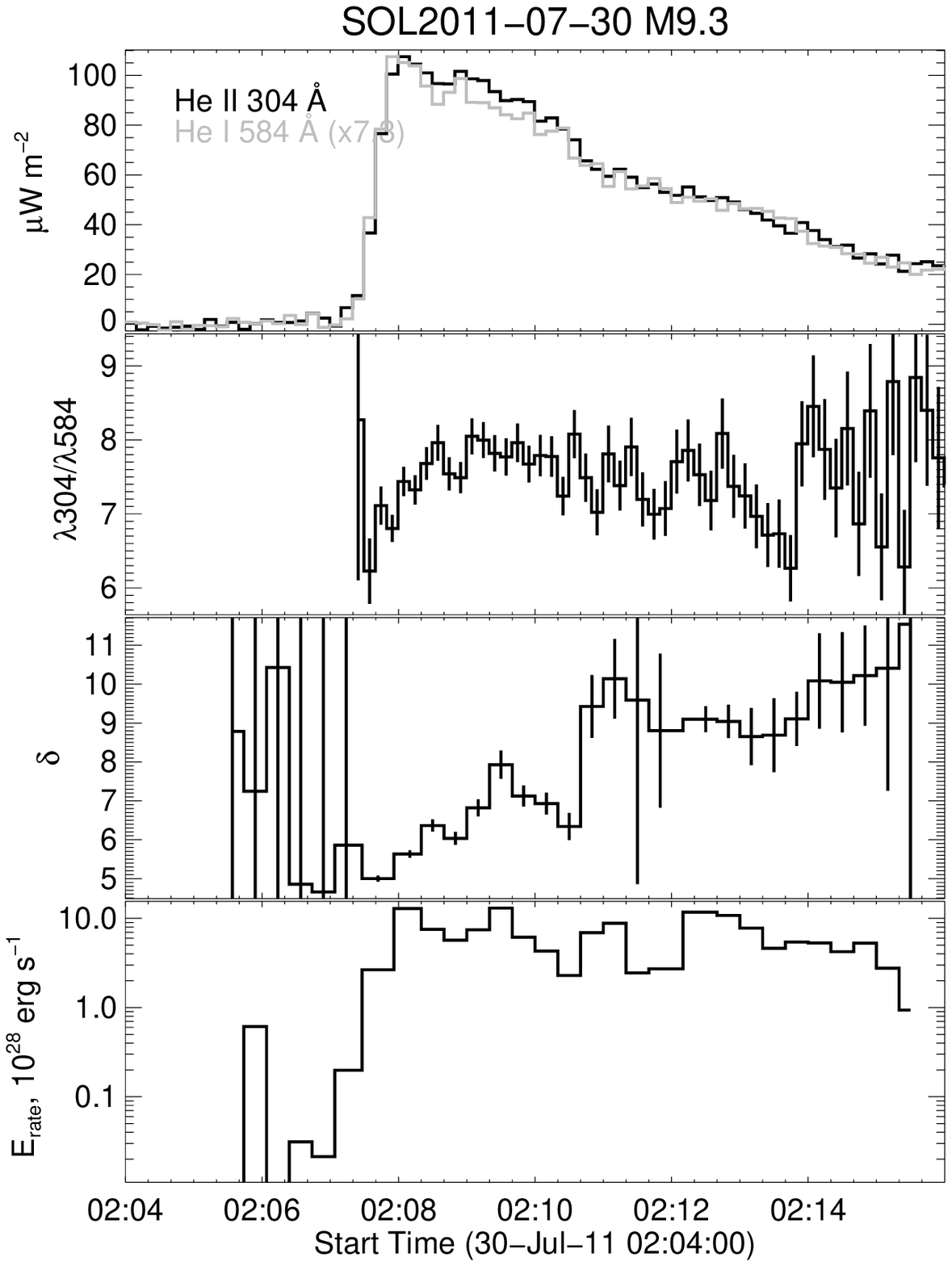}{\figspath/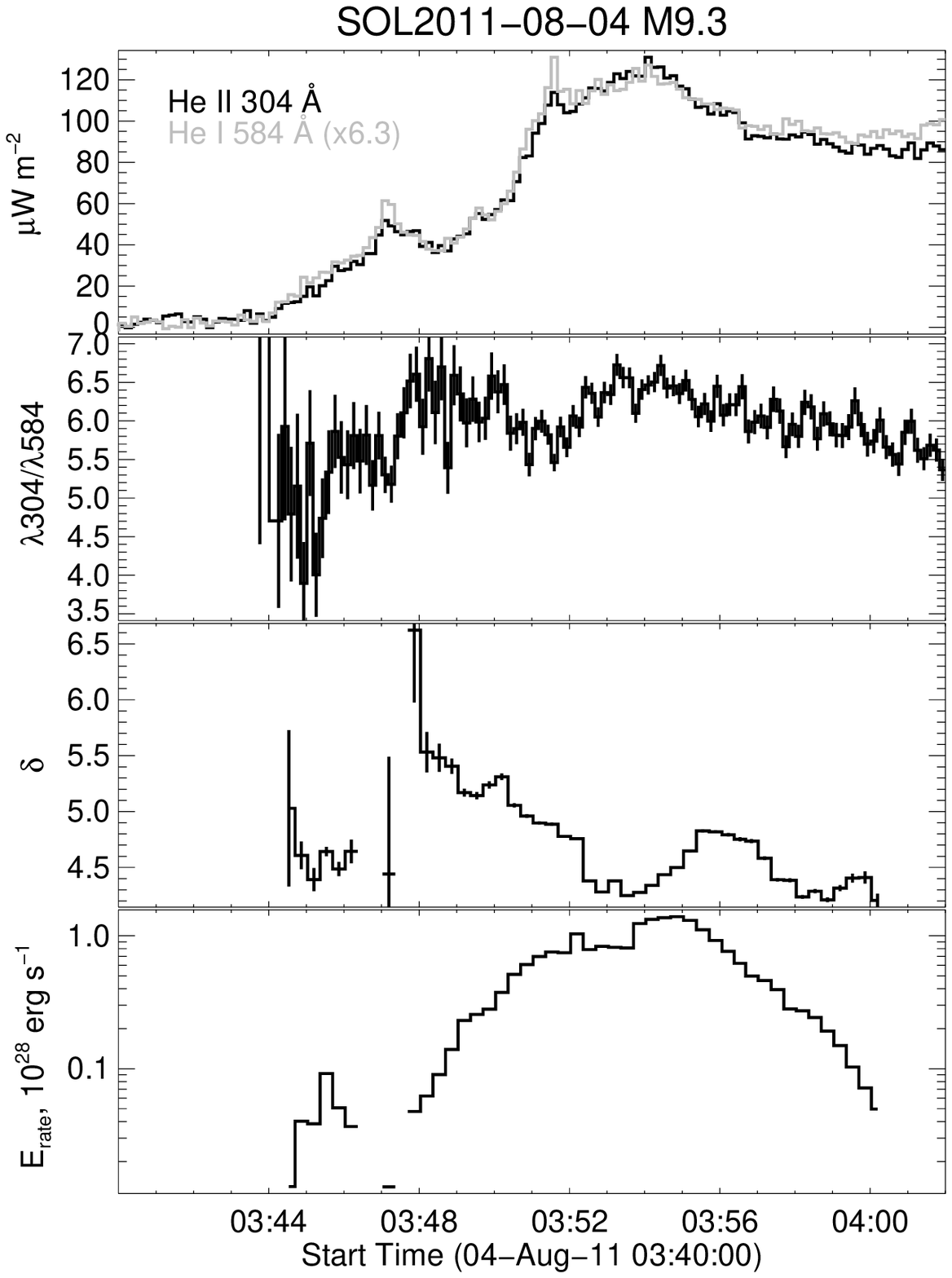}{\figspath/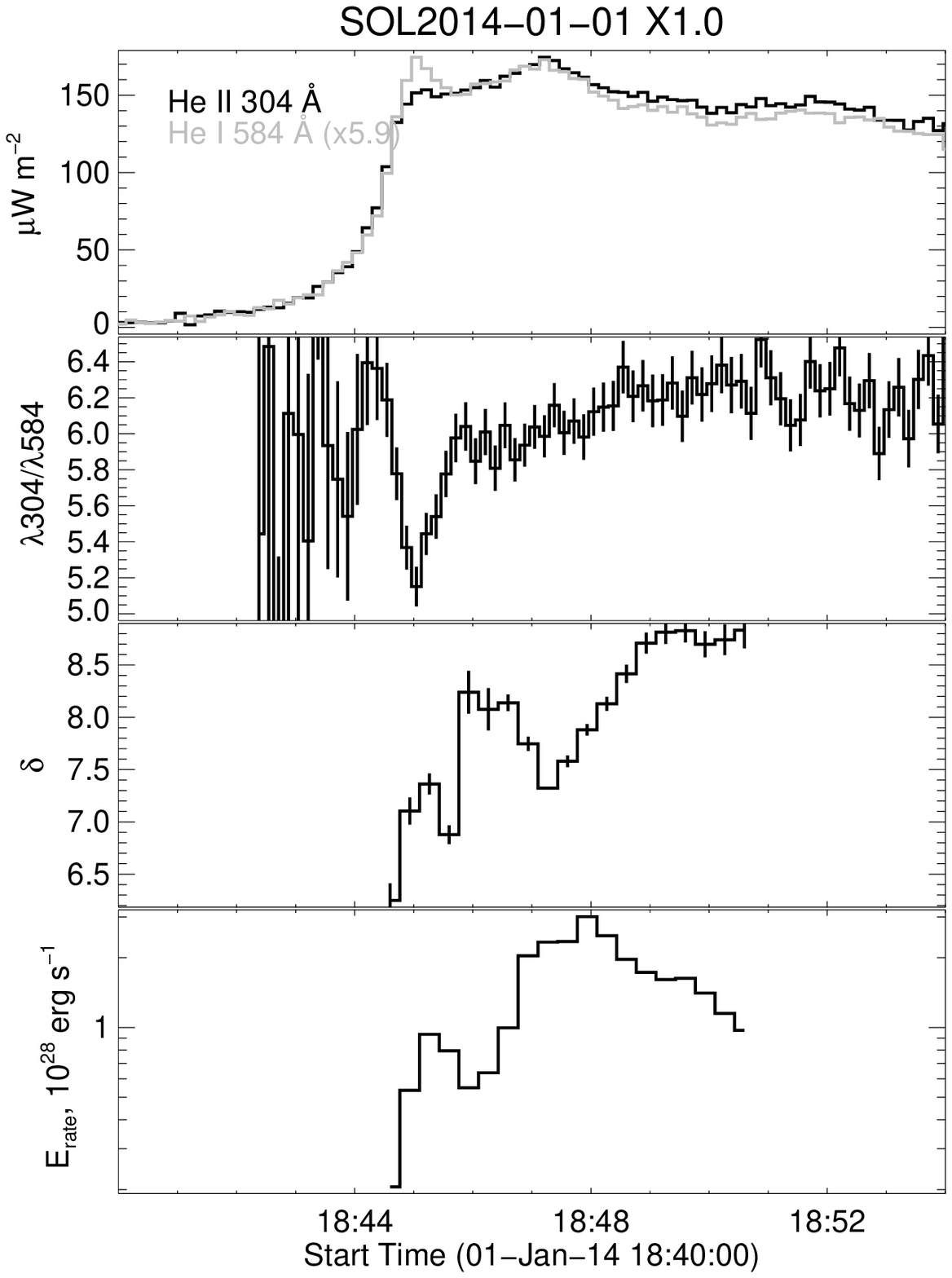}{figobs}{He {\sc ii} 304 \AA\ and He {\sc i} 584 \AA\ irradiances obtained by SDO/EVE, ratio $\lambda$304/$\lambda$584, spectral index of electrons and total electron energy inferred from HXR observations, for four flare events: \emph{Top-left:} SOL2011-02-15, \emph{Top-right:} SOL2011-07-30, \emph{Bottom-left:} SOL2011-08-04, and \emph{Bottom-right:} SOL2014-01-01.}

From Figure \ref{figobs}, although it is not possible to make a direct association between the evolution of ratios $R$ and the beam parameters $\delta$ or $E_\mathrm{rate}$, by comparing the data from the different flares it is clear that the ratio $R$ is larger for higher $E_\mathrm{rate}$ values (our observational proxy for the heating rate), as predicted by our modelling results. Also, the observed ratios $R$ are comparable to the values obtained in the simulations. We now add a few comments regarding each flare. \emph{SOL2011-02-15:} this event has the highest ratio $R$ observed in our sample, beginning with a value around 8 at the onset of the flare and then reaching a plateau around 12 throughout the impulsive phase. It also has the highest $E_\mathrm{rate}$ reaching $\approx 20 \times 10^{28}$ erg s$^{-1}$. \emph{SOL2011-07-30:} $E_\mathrm{rate}$ rises steeply to values $\approx10 \times 10^{28}$ erg s$^{-1}$, with a similar fast change in the ratio $R$ from $6.5 \sim 7.0$ to $7.5 \sim 8$, remaining at this level for the duration of the impulsive phase. \emph{SOL2011-08-04:} $R$ changes from $\approx 5$ at the beginning, when $E_\mathrm{rate}$ is low ($E_\mathrm{rate} < 0.1 \times 10^{28}$ erg s$^{-1}$), to $R \approx 6.5$ when $E_\mathrm{rate} \approx 10^{28}$ erg s$^{-1}$.  \emph{SOL2014-01-01:} after a noisy onset, the ratio goes from $R \approx 5$ (around 18:45~UT) when $E_\mathrm{rate}$ is just below $\approx 10^{28}$ erg s$^{-1}$; $E_\mathrm{rate}$ then rises to about $3 \times 10^{28}$ erg s$^{-1}$, with $R$ reaching values around 6, and maintaining these values for the rest of the impulsive phase. 

We note that the spectral indices $\delta$ vary during these flares, but no variation in $R$ is noticeable at the same timescales. In fact, we did not expect to detect a direct correlation between the beam parameters and the observed $R$ because our results from the simulations show that $R$ is more strongly affected by the energy flux $F$ than by the spectral index $\delta$. In addition, both EVE and HXR observations are spatially integrated, averaging the emission from the different spatial structures present in a flare. If we take into account that different parts of the flare ribbons have their own time evolution with both heating and cooling phases, as noted by \cite{2015ApJ...807L..22G}, the spatially integrated emission observed by EVE and RHESSI or Fermi/GBM will be also a time average of such heating and cooling episodes, potentially smoothing out the variations in $R$ caused by changes in $\delta$.

\section{Conclusions}

We present our results of radiative hydrodynamics simulations of solar flares that indicate an association between the He {\sc ii} 304 \AA\ line intensity with the heating rate in the chromosphere, and that the intensity of He {\sc i} 584 \AA\ is less sensitive to this heating. Under the typical assumption that the flaring chromosphere is collisionally heated by a beam of accelerated electrons, the heating rate $Q$ is controlled by the beam parameters (energy flux $F$, low energy cutoff $E_c$ and spectral index $\delta$). Using RADYN simulations for different electron beam parameters we show that for larger values of $F$ the maximum values of the intensity ratio $R$ of He {\sc ii} to He {\sc i} are larger. We found a similar trend in observations of four flares, using SDO/EVE He irradiance ratios and estimates of the electron beam energy rate obtained from hard X-ray data by RHESSI and Fermi/GBM.

\clearpage 


\acknowledgements The research leading to these results has received funding from the European Community's Seventh Framework Programme (FP7/2007-2013) under grant agreement no. 606862 (F-CHROMA). G.S.K. acknowledges support from a PhD studentship provided by the College of Science and Engineering, University of Glasgow. We would like to thank A.~F.~Kowalski and J.~C.~Allred for providing the RADYN models F9 and F11 used in this analysis.

\bibliography{simoes}  

\end{document}